\documentclass[twocolumn,aps,prl,superscriptaddress,showpacs,tightenlines]{revtex4-1}
\usepackage{mathptmx}
\DeclareMathAlphabet{\mathcal}{OMS}{cmsy}{m}{n}
\usepackage{mathrsfs}

\usepackage[T1]{fontenc}
\usepackage[latin9]{inputenc}
\usepackage{amstext}
\usepackage{amsmath}
\usepackage{graphicx}
\usepackage{esint}
\usepackage{color}
\usepackage[colorlinks, linkcolor=blue, citecolor=blue,hyperindex,bookmarks=false,pdfstartview=FitH]{hyperref}
\makeatletter

\@ifundefined{textcolor}{}
{%
 \definecolor{BLACK}{gray}{0}
\definecolor{WHITE}{gray}{1}
 \definecolor{RED}{rgb}{1,0,0}
 \definecolor{GREEN}{rgb}{0,1,0}
 \definecolor{BLUE}{rgb}{0,0,1}
 \definecolor{CYAN}{cmyk}{1,0,0,0}
 \definecolor{MAGENTA}{cmyk}{0,1,0,0}
 \definecolor{YELLOW}{cmyk}{0,0,1,0}
}

\usepackage{soul}

\makeatother

\begin{document}
\title{Superradiant Phase Transition in the Strong Coupling Regime}

\author{Jin-Feng Huang}
\email{Corresponding author: jfhuang@hunnu.edu.cn}
\affiliation{Key Laboratory of Low-Dimensional Quantum Structures and Quantum Control of Ministry of Education, Key Laboratory for Matter Microstructure and Function of Hunan Province, Department of Physics and Synergetic Innovation Center for Quantum Effects and Applications, Hunan Normal University, Changsha 410081, China}
\affiliation{School of Natural Sciences, University of California, Merced, California 95343, USA}

\author{Lin Tian}
\email{Corresponding author: ltian@ucmerced.edu}
\affiliation{School of Natural Sciences, University of California, Merced, California 95343, USA}

\date{\today}

\begin{abstract}
The Dicke model can exhibit quantum phase transition between the normal and the superradiant phases when the strength of the light-matter coupling exceeds the ultrastrong coupling regime. However, it is challenging to observe this phase transition in practical systems due to limited coupling strength or finite two-photon $A^{2}$ terms. Here we show that by applying a periodic modulation to the frequency of the two-level systems in a standard Dicke model in the strong coupling regime, an anisotropic Dicke model with tunable rotating and counter-rotating terms in the ultrastrong coupling regime can be achieved. We calculate the ground state and the excitation spectrum of this model in terms of the modulation parameters. Our result shows that the superradiant phases can be observed in cavity- or circuit-quantum electrodynamics systems with strong coupling.
\end{abstract}
\maketitle

The Dicke model~\cite{Dicke1954} describes a cavity mode coupled to multiple quantum two-level systems (or qubits) in cavity- or  circuit-quantum electrodynamics (QED) systems~\cite{Carmichael2007PRA, Esslinger2010Nat,Domokos2010PRL, Esslinger2011PRL, FinkPRL2009, MachaNatComm2014, Ciuti2014PRL}. It has been widely studied to exhibit the superradiant phase transition at a critical temperature or a critical light-matter coupling strength~\cite{Brandes2003PRL,Brandes2003PRE, Sachdev1999,Huang2009,Lieb1973AP,Hioe1973PRA}, where the superradiant phase is characterized by macroscopic excitations of the cavity and the qubits. At zero temperature, quantum phase transition (QPT) between the normal and the superradiant phases can occur when the coupling strength exceeds the ultrastrong coupling regime. Recently, dynamical phase transition has been studied both theoretically and experimentally in dissipative Dicke models~\cite{Keeling2012PRA,Tureci2013PRL,Strunz2020PRL,Rey2021PRR, Hemmerich2015PNAS, Esslinger2013RMP, Thompson2020N}.

Despite intensive efforts with both atomic systems and solid-state devices, it is still challenging to observe the ground-state superradiant phase transition in the Dicke model due to limited coupling strength or the two-photon $A^2$ term in practical systems. In particular, for atomic systems in the ultrastrong coupling regime, the $A^2$ terms resulted from second-order effects of the light-matter interaction can prevent the occurrence of the superradiant phase transition~\cite{Zakowicz1975PRL,Ciuti2010NC}. In the superconducting circuit-QED systems, although the ultrastrong coupling regime can now be reached~\cite{Gross2010NP,Solano2010PRL,Semba2017NP,Lupascu2017NP,Steele2017npjQI}, imperfection of the quantum circuits prevents the observation of such phase transition in this parameter regime.

In this Letter, we present an approach that enables the observation of the quantum superradiant phase transition in a standard Dicke model in the strong coupling regime with the strength of the collective light-matter coupling much smaller than qubit and cavity frequencies.
In our approach, by applying a periodic modulation to the frequency of the qubits in the standard Dicke model, a tunable anisotropic Dicke model with ultrastrong rotating and counter-rotating couplings can be generated. The qubit frequency modulation generates sidebands in the energy spectrum of the qubits. By controlling the frequency and magnitude of the modulation, the coupling strengths of the dominate rotating and counter-rotating sidebands with regards to the effective frequencies of the qubits and the cavity in the anisotropic Dicke model can be tuned in a broad range. Both the rotating and the counter-rotating terms can reach the ultrastrong coupling regime. With our parameters, the two-photon $A^2$ terms are far off resonance and can be neglected. This makes it possible to observe the superradiant phase in the cavity-QED setup, which was considered impossible in previous works~\cite{Zakowicz1975PRL,Ciuti2010NC}.
We calculate the ground-state phases and excitation spectra vs the modulation parameters. Our result shows that ground state superradiant phases can be observed in the cavity- or circuit-QED systems in the strong coupling regime. Given the tunability of the effective model in this approach, it can be utilized to study phase transitions in related models such as the Tavis-Cummings model~\cite{TC1968PR} and the Lipkin-Meshkov-Glick model~\cite{LMG1965}. Our work can inspire future studies on implementing quantum phase transitions in engineered quantum systems.

\textbf{Model.} Consider a standard Dicke model, where a cavity mode is coupled to $N$ qubits with the frequency of the qubits periodically modulated. The total Hamiltonian has the form $H_{\rm t}= H_{\rm SD}+H_{A^{2}}+H_{\rm M}(t)$, which includes the Hamiltonian of the standard Dicke model ($\hbar\equiv1$)
\begin{equation}
H_{\rm SD} = \omega_{0}J_{z}+\omega_{c}a^{\dagger}a+g_{0}N^{-1/2}\left(J_{+}+J_{-}\right)(a+a^{\dagger}), \label{eq:H_sd}
\end{equation}
a two-photon $A^{2}$ term $H_{A^{2}}=g_{A^{2}}(a+a^{\dagger})^{2}$ with amplitude $g_{A^{2}}$, and a periodic modulation of the qubit frequency $H_{\textrm{M}}\left(t\right)=\xi\nu\cos\left(\nu t\right)J_{z}$ with dimensionless driving magnitude $\xi$ and modulation frequency $\nu$.
Here $J_{z}=\sum_{i=1}^{N}\sigma_{z}^{\left(i\right)} /2 $ and $J_{\pm}=\sum_{i=1}^{N}\sigma_{\pm}^{\left(i\right)}$ are collective spin operators defined as the sum of the Pauli operators $\sigma_{z,\pm}^{\left(i\right)}$ of the qubits, $ \omega_{0}$ is the energy splitting of the qubits, $a$ ($a^{\dagger}$) is the annihilation (creation) operator and $\omega_{c}$ is the frequency of the cavity mode, and $g_{0}/\sqrt{N}$ is the coupling strength between an individual qubit and the cavity mode. The collective spin operators obey the usual angular momentum commutation relations $\left[J_{z},J_{\pm}\right]=\pm J_{\pm}$ and $\left[J_{+},J_{-}\right]=2J_{z}$, and have the angular momentum eigenstates $\vert j,m\rangle$ with maximum eigenvalue $j_{\rm max}=N/2$ and $m\in [-j, \, j]$.

The standard Dicke model can exhibit quantum phase transition from a normal phase to a superradiant phase with macroscopic cavity displacement and qubit excitations. This phase transition occurs when the collective coupling strength reaches the ultrastrong coupling regime with $g_{0}\ge\sqrt{\omega_{0}\omega_{c}}/2$~\cite{Dicke1954,Huang2009}. The amplitude of the two-photon Hamiltonian $H_{A^{2}}$ can be written as $g_{A^{2}}=\chi g_{0}^{2}/\omega_{0}$ with $\chi$ being a dimensionless coefficient.  In cavity QED, governed by the Thomas-Reiche-Kuhn sum rule, $\chi\geq1$, which prevents the occurrence of the superradiant phase transition~\cite{Ciuti2010NC}. Even though we can have $\chi\ll 1$ in circuit QED, other factors such as the parameter spread of the qubits in the ultrastrong coupling regime could prevent the observation of this phase transition.

Below we derive the effective Hamiltonian of the standard Dicke model under periodic modulation. Let $H_{0}^{(1)}= \omega_{0}J_{z}+\omega_{c}'a^{\dagger}a+H_{\rm M}(t)$ with $\omega_{c}'=\omega_{c}+2g_{A^{2}}$, which includes the modulation of the qubit frequency. In the rotating frame of $H_{0}^{(1)}$, the effective Hamiltonian of the modulated Dicke model is $H_{\rm rot}^{(1)}=V_{1}^{\dagger}(t) (H_{\rm t}-H_{0}^{(1)})V_{1}(t)$ with $V_{1}(t)=\exp[-i\int_{0}^{t}H_{0}^{(1)}(\tau)d\tau]$. After omitting the constant term in $H_{A^{2}}$, we find that
\begin{align}
H_{\rm rot}^{(1)} = & \frac{g_{0}}{\sqrt{N}}  \sum_{n=-\infty}^{\infty} J_{n}\left(\xi\right)\left[J_{+}( ae^{i \delta_{n}t}+a^{\dagger}e^{i\Delta_{n}t})+\mbox{H.c.}\right]\nonumber \\
 &  +g_{A^{2}}(a^{2}e^{-2i\omega_{c}'t}+a^{\dagger2}e^{2i\omega_{c}'t}),\label{eq:Hrot1}
\end{align}
where $\delta_{n}=\omega_{0}-\omega_{c}'+n\nu$, $\Delta_{n}=\omega_{0}+\omega_{c}'+n\nu$, and $J_{n}$ is the $n$th Bessel function of the first kind with integer number $n$. Here we have used the Jacobi\textendash Anger identity: $\exp[i\xi\sin\left(\nu t\right)] = \sum_{n=-\infty}^{\infty}J_{n}\left(\xi\right)\exp(in\nu t)$ for Bessel functions. As shown in Eq.~(\ref{eq:Hrot1}), the modulation of the qubit frequency generates spectral sidebands in the rotating (counter-rotating) terms with detuning $\delta_{n}$ ($\Delta_{n}$) and coupling amplitude $(g_{0}/\sqrt{N}) J_{n}\left(\xi\right)$. The sidebands are separated by the modulation frequency $\nu$. The amplitudes of the sidebands can be adjusted by varying the dimensionless modulation amplitude $\xi$.

We introduce a second rotating frame defined by the Hamiltonian $H_{0}^{(2)}= -\tilde{\omega}_{0}J_{z} - \tilde{\omega}_{c}a^{\dagger}a$ with the effective qubit and cavity frequencies $\tilde{\omega}_{0}=\left(\delta_{n_{0}} + \Delta_{m_{0}}\right) /2$ and $\tilde{\omega}_{c}=\left(\Delta_{m_{0}}-\delta_{n_{0}}\right) /2$, respectively. Here by choosing appropriate qubit frequency and modulation frequency, we can select a rotating sideband $n_{0}$ and a counter-rotating sideband $m_{0}$, where the effective coupling $\lambda_{\rm r}=g_{0}J_{n_{0}}\left(\xi\right)$ [$\lambda_{\rm cr}=g_{0}J_{m_{0}}\left(\xi\right)$] for the sideband can reach the ultrastrong coupling regime with respect to its rotating frequency $\vert \delta_{n_{0}} \vert$ ($\vert \Delta_{m_{0}} \vert$).
Meanwhile, under the condition $\omega_{0}, \omega_{c}, \nu \gg g_{0}$, all other sidebands are fast rotating, i.e.,
\begin{equation}
g_{0} \vert J_{n}(\xi)\vert,\,\, g_{0} \vert J_{m}(\xi)\vert \ll \nu, \vert \delta_{n\ne n_{0}}\vert, \vert \Delta_{m\ne m_{0}}\vert. \label{eq:othersidebands}
\end{equation}
With $g_{A^{2}}\ll 2\omega_{c}^{\prime}$, the $a^2$ and $a^{\dag2}$ terms with oscillating frequencies $\pm 2\omega_{c}'$ in $H_{A^{2}}$ are also fast rotating~\cite{supple}. In the rotating frame of $H_{0}^{(2)}$, the effective Hamiltonian is $H_{\rm rot}^{(2)}=V_{2}^{\dagger}\left(t\right)( H_{\rm rot}^{(1)} - H_{0}^{(2)} )V_{2}\left(t\right)$ with the unitary transformation $V_{2}\left(t\right)=\exp[-i H_{0}^{(2)}t]$. Under the rotating-wave approximation with all fast-rotating sidebands with $n\ne n_{0}$ and $m\ne m_{0}$ neglected, the Hamiltonian becomes
\begin{equation}
H_{\rm rot}^{(2)} = \tilde{\omega}_{0}J_{z} + \tilde{\omega}_{c}a^{\dagger}a +N^{-1/2}\left[ J_{+}\left(\lambda_{\textrm{r}}a+\lambda_{\textrm{cr}}a^{\dagger}\right)+\textrm{H.c.}\right].\label{eq:Hrot2A}
\end{equation}
This Hamiltonian describes an anisotropic Dicke Hamiltonian with effective couplings $\lambda_{\textrm{r}}$ and $\lambda_{\textrm{cr}}$ for the rotating and the counter-rotating terms, respectively.

\begin{figure}
\includegraphics[bb=0bp 0bp 399bp 318bp,clip,scale=0.6]{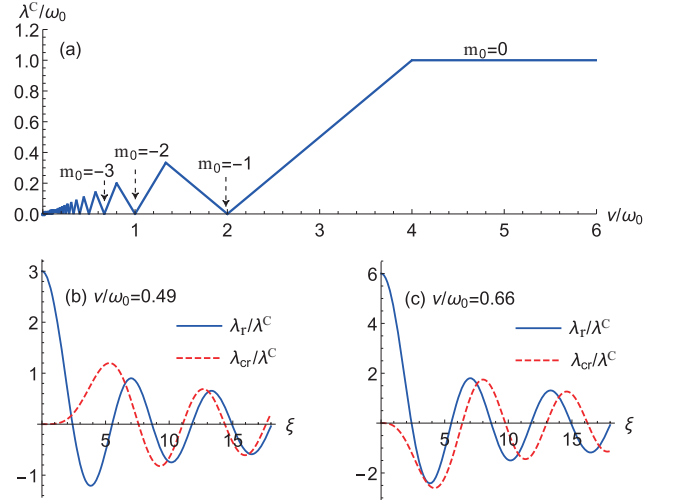}
\caption{(a) The critical coupling $\lambda^{C}/\omega_{0}$ vs the modulation frequency $\nu/\omega_{0}$.
(b) and (c) The ratios $\lambda_{\textrm{r}}/\lambda^{C}$ and $\lambda_{\textrm{cr}}/\lambda^{C}$ vs the driving amplitude $\xi$ for $g_{0}/\omega_{0}=0.06$ at $\nu/\omega_{0}=0.49$ and $0.66$, respectively. Here we choose $\omega_{0}=\omega_{c}'$.}
\label{fig1}
\end{figure}
\textbf{Ultrastrong coupling.} The parameters in Eq.~(\ref{eq:Hrot2A}) can be adjusted to reach the ultrastrong coupling regime by varying the frequency $\nu$ and the amplitude $\xi$ of the periodic modulation. We define $\lambda^{C}=\sqrt{\tilde{\omega}_{0}\tilde{\omega}_{c}}$ as the critical coupling, which is crucial for our discussion of the superradiant phase transition and only depends on the driving frequency $\nu$.
In Fig.~\ref{fig1}(a), we plot $\lambda^{C}/\omega_{0}$ vs $\nu/\omega_{0}$.
At $\omega_{0}=\omega_{c}'$, the index for the rotating sideband is $n_{0}=0$ with $\delta_{n_{0}}=0$. Whereas the index $m_{0}$ for the counter-rotating sideband varies with
$\nu/\omega_{0}$ and reaches $m_{0}=0$ when $\nu>2\left(\omega_{0}+\text{\ensuremath{\omega_{c}'}}\right)$.
Each V-shaped valley shares the same $m_{0}$. At $\nu_{1}=-2\omega_{0}/{m_{0}}$ and $\nu_{2}=-2\omega_{c}'/{m_{0}}$, $\lambda^{C}=0$. At $\omega_{0}=\omega_{c}'$, $\vert \nu_{1} -\nu_{2}\vert = 0$.

The effective coupling strengths depend on the driving amplitude $\xi$ in the form of the Bessel functions. In Figs.~\ref{fig1}(b, c), we plot $\lambda_{\textrm{r}}/\lambda^{\mathrm{C}}$ and $\lambda_{\textrm{cr}}/\lambda^{\mathrm{C}}$ vs $\xi$ at $g_{0}=0.06\omega_{0}$, $\omega_{0}=\omega_{c}'$ for two values of the driving frequency $\nu$.
For $\nu/\omega_{0}=0.49$, $n_{0}=0$ and $m_{0}=-4$. As $\xi$ increases, $\lambda_{\mathrm{r}}/\lambda^{\mathrm{C}}$ oscillates smoothly between $3$  and $-1.208$ with reducing amplitude, and $\lambda_{\mathrm{cr}}/\lambda^{\mathrm{C}}$ oscillates between
$1.199$ and $-0.823$. Both couplings can be tuned to zero when the corresponding Bessel function becomes zero.
For $\nu/\omega_{0}=0.66$, $n_{0}=0$ and $m_{0}=-3$, and $\lambda_{\textrm{r}}$ and $\lambda_{\textrm{cr}}$ exhibit similar behavior.
This result shows that both $\lambda_{\textrm{r}}$ and $\lambda_{\textrm{cr}}$ can be tuned in a broad range and can enter the ultrastrong coupling regime. In particular, in the neighborhood of the valley dips in Fig.~\ref{fig1}(a), these couplings can exceed the magnitude of the critical coupling $\lambda^{C}$. Our system can hence demonstrate rich quantum phenomena as discussed below.

\textbf{Quantum phase transition.}
In the Holstein-Primakoff representation, the collective angular momentum operators can be written as $J_{+}=b^{\dagger}\sqrt{N-b^{\dagger}b}$, $J_{-}=\sqrt{N-b^{\dagger}b}\ b$, and $J_{z}=b^{\dagger}b-N/{2}$~\cite{HP 1940} in terms of a bosonic mode with annihilation (creation) operator $b$ ($b^{\dagger}$) and $\left[b,\,b^{\dagger}\right]=1$. The anisotropic Dicke model in (\ref{eq:Hrot2A}) then becomes:
\begin{eqnarray}
H_{\rm rot}^{(2)} &=& \tilde{\omega}_{0} b^{\dagger} b + \tilde{\omega}_{c} a^{\dagger} a - N\tilde{\omega}_{0}/2 \nonumber \\ && +[ b^{\dagger}\sqrt{1-b^{\dagger} b /N}  (\lambda_{\textrm{r}}a+\lambda_{\textrm{cr}}a^{\dagger})+\textrm{H.c.}].\label{eq:Hrot2}
\end{eqnarray}
The ground state of the anisotropic Dicke model can be either in a normal phase with $\langle a\rangle =\langle b\rangle =0$ or in a superradiant phase with finite $\langle a\rangle$ and $\langle b\rangle$, depending on the coupling strengths $\lambda_{\textrm{r}}$ and $\lambda_{\textrm{cr}}$.
To derive the ground state, we use a mean-field approach~\cite{Huang2009,Ciuti2014PRL,Brandes2003PRL,Brandes2003PRE} and write the bosonic operators as $a\rightarrow c+\alpha$ and $b\rightarrow d+\beta$,
where $\alpha=\left\langle a\right\rangle $ ($\beta=\left\langle b\right\rangle $) is the semiclassical displacement of the cavity (collective qubit mode), and operator $c$ ($d$) represents the quantum fluctuation of the displaced cavity (qubit) mode with $\left\langle c\right\rangle=0$ ($\left\langle d\right\rangle=0$). Denote $\vec{v}_{S}=\left(c, d, c^{\dagger}, d^{\dagger}\right)^{T}$. In the thermodynamic limit $N\rightarrow\infty$, using a Taylor expansion of the Hamiltonian $H_{\rm rot}^{(2)}$ in terms of the fluctuation operators $c$ and $d$ and keeping to the second order terms, we have $H_{\rm rot}^{(2)}=H_{II} + H_{I}+E_{\textrm{G}}$, where $H_{II}=\vec{v}_{S}^{\dagger} G\vec{v}_{S}$ with $G$ being a $4\times 4 $ Hermitian matrix, $H_{I}=\vec{\Omega}^{T}\vec{v}_{S}$ with $\vec{\Omega}$ being a $4\times 1 $ vector, and $E_{\textrm{G}}$ is a constant. Here the matrix $G$, vector $\vec{\Omega}$, and $E_{\textrm{G}}$ all depend on the displacements $\alpha$ and $\beta$, details of which can be found in \cite{supple}. When $\alpha$ and $\beta$ correspond to the ground state displacements, the linear term disappears with $H_{I}=0$. Hence by solving the equation $\vec{\Omega}=0$, we can find the solution to the semiclassical displacements $\alpha$ and $\beta$ and derive the ground state energy $E_{\textrm{G}}$ and the matrix $G$.

\begin{figure}
\includegraphics[bb=0bp 29bp 584bp 271bp,clip,scale=0.43]{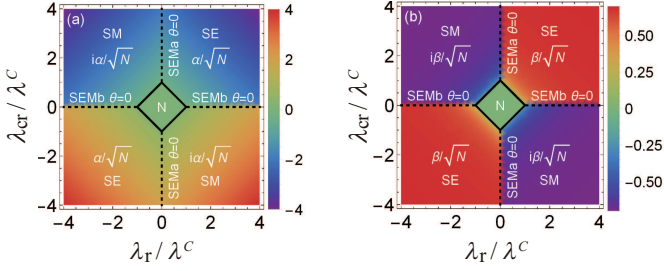}
\caption{The displacements (a) $\alpha/\sqrt{N}$ (or $i\alpha/\sqrt{N}$) and (b) $\beta/\sqrt{N}$ (or $i\beta/\sqrt{N}$) versus the couplings $\lambda_{\textrm{r}}/\lambda^{C}$ and $\lambda_{\textrm{cr}}/\lambda^{C}$. In the SEMa and SEMb phases, we choose $\theta=0$. Other parameters are $\nu/\omega_{0}=0.49$, $g_{0}/\omega_{0}=0.06$, and $\omega_{0}=\omega_{c}'$.}
\label{fig2}
\end{figure}
The displacements of the cavity and qubit modes in the anisotropic Dicke model are plotted in Figs.~\ref{fig2}(a, b). When $\left|\lambda_{\textrm{cr}}\pm\lambda_{\textrm{r}}\right|<\lambda^{C}$, $\alpha=\beta=0$, which corresponds to the normal phase as labelled by N in Fig.~\ref{fig2}. Outside the normal phase, when $\lambda_{\textrm{r}}\lambda_{\textrm{cr}}>0$, the ground state  is in the superradiant electric (SE) phase with two sets of solutions $\left(\alpha, \beta\right )=\pm\left(\alpha_{0}, \beta_{0}\right )$ and $\alpha_{0},\beta_{0}$ being real numbers.
When $\lambda_{\textrm{r}}\lambda_{\textrm{cr}}<0$, the ground state is in the superradiant magnetic (SM) phase with two sets of imaginary number displacements $\pm \left(i\alpha_{0},  i\beta_{0}\right)$.
Along the $y$-axis when $\vert \lambda_{\textrm{cr}}\vert >\lambda^{C}$, the system is in the superradiant electromagnetic a (SEMa) phase, where the semiclassical displacements are $\left( \alpha_{0} e^{-i\theta},  \beta_{0} e^{i\theta}\right )$ with an arbitrary but opposite phase factor $\theta$. Similarly, along the $x$-axis when $\vert \lambda_{\textrm{r}}\vert >\lambda^{C}$, the system is in the superradiant electromagnetic b (SEMb) phase with displacements $\left( \alpha_{0} e^{i\theta}, \beta_{0} e^{i\theta}\right)$. Details of these solutions can be found in \cite{supple}.
The solutions in the SE and SM phases break the $Z_{2}$ symmetry of the model when $\lambda_{\textrm{r}}\lambda_{\textrm{cr}}\ne 0$, whereas the solutions in the SEMa and SEMb phases break the U(1) symmetry of the model when $\lambda_{\textrm{r}}\lambda_{\textrm{cr}}= 0$.
Note that for $\lambda_{\textrm{r}}=\lambda_{\textrm{cr}}$, the condition for the normal phase becomes $\vert \lambda_{\textrm{r}}\vert<\lambda^{C}/2$, which agrees with the result for a standard Dicke model~\cite{Brandes2003PRE}. The solid and the dashed lines in Fig.~\ref{fig2} indicate the phase boundaries separating these phases.

\begin{figure}
\includegraphics[bb=0bp 0bp 543bp 524bp, clip,scale=0.32]{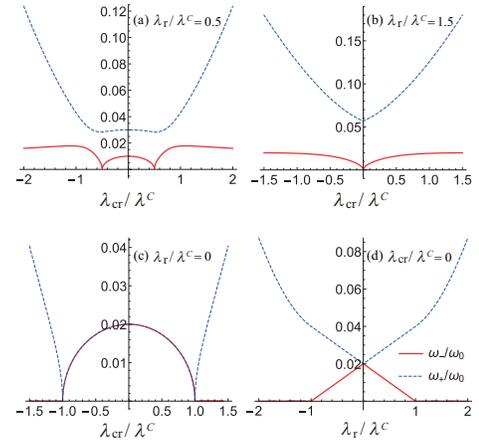}
\caption{(a-c) Quasiparticle spectrum $\omega_{\pm}$ vs the relative coupling  $\lambda_{\textrm{cr}}/\lambda^{C}$ at $\lambda_{\textrm{r}}/\lambda^{C}=0.5, 1.5, 0$, respectively. (d) $\omega_{\pm}$ vs the relative coupling $\lambda_{\textrm{r}}/\lambda^{C}$ at $\lambda_{\textrm{cr}}/\lambda^{C}=0$. Other parameters are the same as those in Fig.~\ref{fig2}.}
\label{fig3}
\end{figure}
Using the Hopfield-Bogoliubov transformation on $H_{II}$~\cite{Hopfield1958}, the system Hamiltonian can be diagonalized as $H_{\rm rot}^{(2)} =\sum_{i=\pm}\omega_{i}^{p}\gamma_{i}^{p\dagger}\gamma_{i}^{p}+E_{\textrm{G}}^{p}$. Here $\gamma_{i}^{p}$ ($\gamma_{i}^{p\dagger}$) is the annihilation (creation) operator of one of the quasiparticles in the ground state phase $p$ with frequency $\omega_{i}^{p}$, and $E_{\textrm{G}}^{p}$ is the ground state energy in phase $p$. The operator $\gamma_{i}^{p}$ is a linear combination of the operators $c, c^{\dagger}, d, d^{\dagger}$ for the superradiant phases and $a, a^{\dagger}, b, b^{\dagger}$ for the normal phase, with the commutation relation $[\gamma_{i}^{p},\,\gamma_{j}^{p\dagger}]=\delta_{ij}$. Details of the quasiparticle spectrum for different phases are given in \cite{supple}.
In Fig.~\ref{fig3}, we plot the quasiparticle spectrum as functions of the coupling $\lambda_{\textrm{cr}}$ or $\lambda_{\textrm{r}}$. The superscript "p" that refers to the specific phase in the quasiparticle frequency is omitted.
For $\lambda_{\mathrm{r}}/\lambda^{C}=0.5$, the critical points occur at $\lambda_{\mathrm{cr}}/\lambda^{C}=\pm0.5$ when $\omega_{-}=0$, as shown in Fig.~\ref{fig3}(a). The system experiences a second order phase transition from the normal phase at $\vert\lambda_{\mathrm{cr}}/\lambda^{C}\vert<0.5$ to the SE or SM phase at $\vert\lambda_{\mathrm{cr}}/\lambda^{C}\vert>0.5$, as shown in Fig.~\ref{fig2}.
For $\lambda_{\mathrm{r}}/\lambda^{C}=1.5$ presented in Fig.~\ref{fig3}(b), a single critical point occurs at $\lambda_{\mathrm{cr}}/\lambda^{C}=0$, which corresponds to a Goldstone mode in the SEMb phase.
For $\lambda_{\mathrm{r}}/\lambda^{C}=0$ along the $y$-axis, the critical points are at $\lambda_{\mathrm{cr}}/\lambda^{C}=\pm1$, corresponding to a phase transition between the normal phase and the SEMa phase, as shown in Fig.~\ref{fig3}(c). Similarly, for $\lambda_{\mathrm{cr}}/\lambda^{C}=0$ along the $x$-axis, the normal-SEMb phase transition occurs at $\lambda_{\mathrm{r}}/\lambda^{C}=\pm1$, as shown in Fig.~\ref{fig3}(d). In both the SEMa and SEMb phases, $\omega_{-}=0$, corresponding to Goldstone excitations resulted from the U(1) symmetry of the model along the $x$ and $y$ axes.

\begin{figure}
\includegraphics[bb=0bp 0bp 417bp 374bp,clip,scale=0.52]{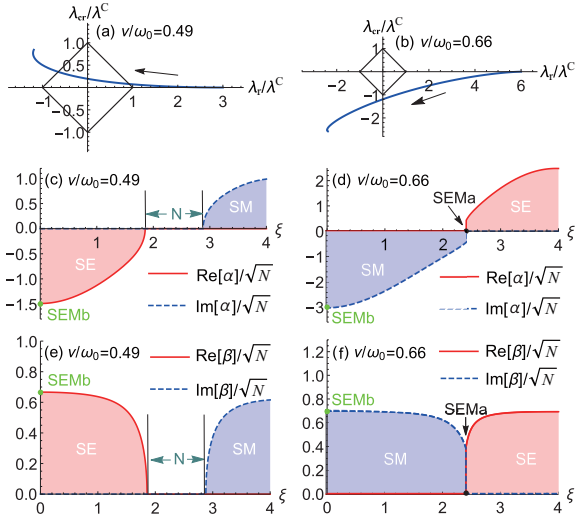}
\caption{(a, b) The trajectory of the effective couplings $\left( \lambda_{\mathrm{r}}/\lambda^{C}, \lambda_{\mathrm{cr}}/\lambda^{C}\right )$ when the driving amplitude $\xi$ increases from 0 to 4 as indicated by the arrows. (c, d) The displacement $\alpha/\sqrt{N}$ and (e, f) the displacement $\beta/\sqrt{N}$ vs the driving amplitude $\xi$. (a), (c) and (e) are for $\nu/\omega_{0}=0.49$. (b), (d) and (f) are for $\nu/\omega_{0}=0.66$. For SEMa and SEMb phases, we set $\theta=0$. Other parameters are the same as those in Figs.~\ref{fig1}(b) and (c).}
\label{fig4}
\end{figure}
\textbf{Manipulation of quantum phases.} By controlling the parameters of the qubit frequency modulation, the effective rotating and counter-rotating couplings in the engineered anisotropic Dicke model can reach the ultrastrong coupling regime with superradiant ground states, even if the physical coupling strength is only in the strong coupling regime. In particular, as shown in Fig.~\ref{fig1}(a), the critical coupling $\lambda^{C}\rightarrow 0$ in the neighborhood of a V-shaped minimum, which results in diminishing normal phase region.

In Figs.~\ref{fig4}(a, b), we plot the trajectories of the effective couplings $\left( \lambda_{\mathrm{r}}/\lambda^{C}, \lambda_{\mathrm{cr}}/\lambda^{C}\right )$ vs the driving amplitude $\xi$ at $\nu/\omega_{0}=0.49$ and $0.66$, respectively. It can be seen that the trajectories evolve through several superradiant phases and the normal phase. In Figs.~\ref{fig4}(c-f), we plot the ground state displacements $\alpha$ and $\beta$ vs $\xi$ for the corresponding values of $\nu/\omega_{0}$. As labelled in the plots, the solid curves are for the real parts and the dashed curves are for the imaginary parts of $\alpha$ and $\beta$. We also indicate the corresponding ground state phases in the plots.
At $\nu/\omega_{0}=0.49$ in Figs.~\ref{fig4}(c, e), the ground state is in the SE phase when $0<\xi<1.856$, where $\lambda_{\textrm{r}}\lambda_{\textrm{cr}}>0$ and $\left|\lambda_{\textrm{r}}+\lambda_{\textrm{cr}}\right|>\lambda^{C}$. In the region $2.880<\xi< 4$, the ground state is in the SM phase with $\lambda_{\textrm{r}}\lambda_{\textrm{cr}}<0$ and $\left|\lambda_{\textrm{r}}-\lambda_{\textrm{cr}}\right|>\lambda^{C}$. The SEMb phase with $\lambda_{\textrm{cr}}=0$ locate at $\xi=0$. The normal phase appears when $1.856<\xi<2.880$. Here the dependence of $\lambda_{\textrm{r}}/\lambda^{C}, \lambda_{\textrm{cr}}/\lambda^{C}$ vs $\xi$ can be seen in Fig.~\ref{fig1}(b).
At $\nu/\omega_{0}=0.66$ with the increase of $\xi$, the ground state is in the SEMb, SM, SEMa, and SE phases sequentially, as shown in Figs.~\ref{fig4}(d, f). The dependence of $\lambda_{\textrm{r}}/\lambda^{C}, \lambda_{\textrm{cr}}/\lambda^{C}$ vs $\xi$ can be found in Fig.~\ref{fig1}(c). With these two values of $\nu/\omega_{0}$,
all normal and superradiant phases can be experienced.

The above result shows that the ground state of the engineered anisotropic Dicke model can be in a superradiant phase when the collective qubit-cavity coupling is only in the strong coupling regime with $g_{0}/\omega_{0}=0.06$. For superconducting qubits with $\omega_{0}/2\pi=10$ GHz, this corresponds to a collective coupling of $g_{0}/2\pi=600$ MHz. For a small array of $N=4$ qubits, the individual qubit-cavity coupling is then $\left( g_{0}/\sqrt{N}\right)/2\pi=300$ MHz, well within the reach of current technology~\cite{squbit_review1, BlaisPRA2004}. In comparison, the ground state of the standard Dicke model can be in the superradiant phase only when $g_{0}/\omega_{0}>0.5$ in the ultrastrong coupling regime. This requires a collective coupling of $g_{0}/2\pi=5$ GHz. Even if the individual qubit-cavity coupling can reach $1$ GHz, it would require an array of $N=25$ qubits to achieve such collective coupling.
Our result hence shows that the normal-superradiant phase transition can be implemented with practical physical systems, such as superconducting qubits and cavity mode in the strong coupling regime. Meanwhile, various superradiant phases such as SE, SM, SEMa and SEMb phases can all be reached by varying the parameters of the frequency modulation. By manipulating the parameters, one can demonstrate rich physics in different superradiant phases, such as the Goldstone modes in the SEMa and SEMb phases.

\textbf{Conclusions.} We studied a scheme that can generate an anisotropic Dicke model with ultrastrong coupling via classical control of engineered quantum systems. By applying properly-designed qubit frequency modulation to a standard Dicke model in the strong coupling regime, the effective rotating and counter-rotating couplings can be tuned in a broad range and reach the ultrastrong coupling regime. We show that various superradiant phases and the normal phase can be achieved in the ground state of this anisotropic Dicke model. Our result demonstrates that superradiant phases can be implemented in practical physical systems with a collective light-matter coupling in the strong coupling regime, and the normal-superradiant phase transition can be observed. With our parameters, the two-photon $A^{2}$ terms that could prevent the implementation of the superradiant phases only have negligible effect on the engineered Hamiltonian.

\begin{acknowledgments}
J.-F.H. is supported in part by the National Natural Science Foundation of China (Grant No. 12075083) and Natural Science Foundation of Hunan Province, China (Grant No. 2020JJ5345). L.T. is supported by the NSF awards No. 2006076 and No. 2037987.
\end{acknowledgments}

\end{document}